\documentclass{rmf-d}
\usepackage{nopageno,rmfbib,multicol,times,epsf,amsmath,amssymb,cite}
\usepackage[latin1]{inputenc}
\usepackage[]{caption2}
\usepackage{graphicx}
\def\rmfcornisa{
\rmf
\hfill June 2022}

\newcommand{\tcite}[1]{~\cite{#1}}
\newcommand{\tref}[1]{~\ref{#1}}

\def\rmfcintilla{{\it Rev.\ Mex.\ Fis.\/} {\bf}}
\clearpage \rmfcaptionstyle 
\begin{document}
\title{Unveiling the proton structure via \\ transverse-momentum-dependent gluon distributions
\vspace{-6pt}}

\author{Alessandro Bacchetta}
\address{Dipartimento di Fisica, Universit\`a di Pavia, via Bassi 6, I-27100 Pavia
\\
INFN Sezione di Pavia, via Bassi 6, I-27100 Pavia, Italy}

\author{Francesco Giovanni Celiberto}
\address{European Centre for Theoretical Studies in Nuclear Physics and Related Areas (ECT*), I-38123 Villazzano, Trento, Italy
\\
Fondazione Bruno Kessler (FBK), I-38123 Povo, Trento, Italy
\\
INFN-TIFPA Trento Institute of Fundamental Physics and Applications, I-38123 Povo, Trento, Italy}

\author{Marco Radici}
\address{INFN Sezione di Pavia, via Bassi 6, I-27100 Pavia, Italy}

\maketitle
\recibido{16 January 2022}{3 March 2022
\vspace{-12pt}}
\begin{abstract}
\vspace{1em} We present exploratory studies of the proton content via unpolarized and polarized transverse-momentum-dependent gluon distributions in the proton at leading twist. We make use of an enhanced spectator-model approach to encode in our densities both small- and moderate-$x$ effects. These studies are relevant to the investigation of the inner structure of hadrons via tomographic analyses at new-generation colliding machines.
\vspace{1em}
\end{abstract}
\keys{proton tomography, gluon content, TMD gluon distributions  \vspace{-4pt}}
\pacs{   \bf{\textit{12.38.-t, 12.38.Aw, 12.39.St, 14.20.Dh }}    \vspace{-4pt}}
\begin{multicols}{2}

\section{Introductory remarks}
\label{intro}

\begin{figure*}[t]
 
 \centering

 \hspace{0.35cm}
 \includegraphics[width=0.45\textwidth]{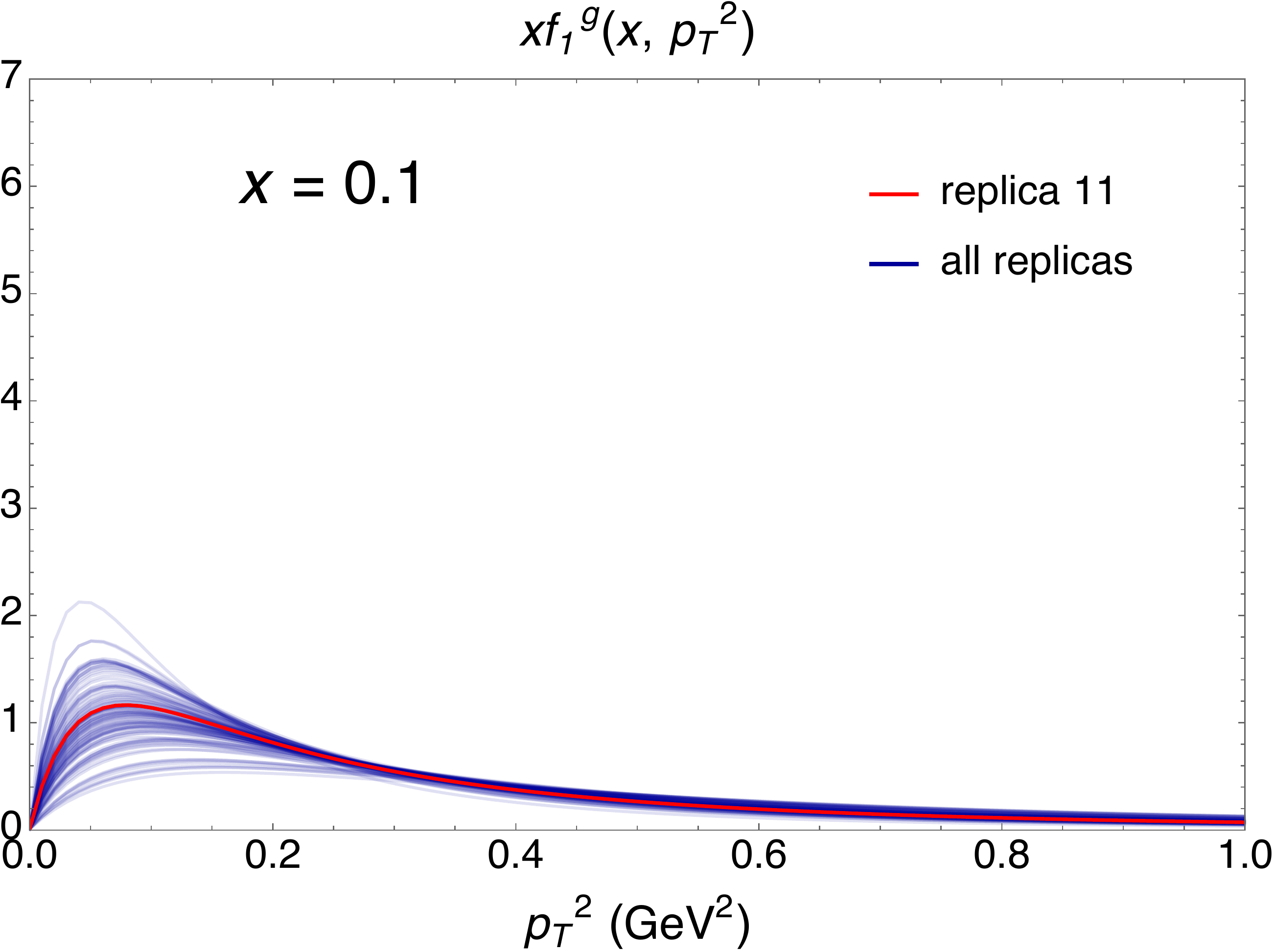}
 \hspace{0.75cm}
 \includegraphics[width=0.45\textwidth]{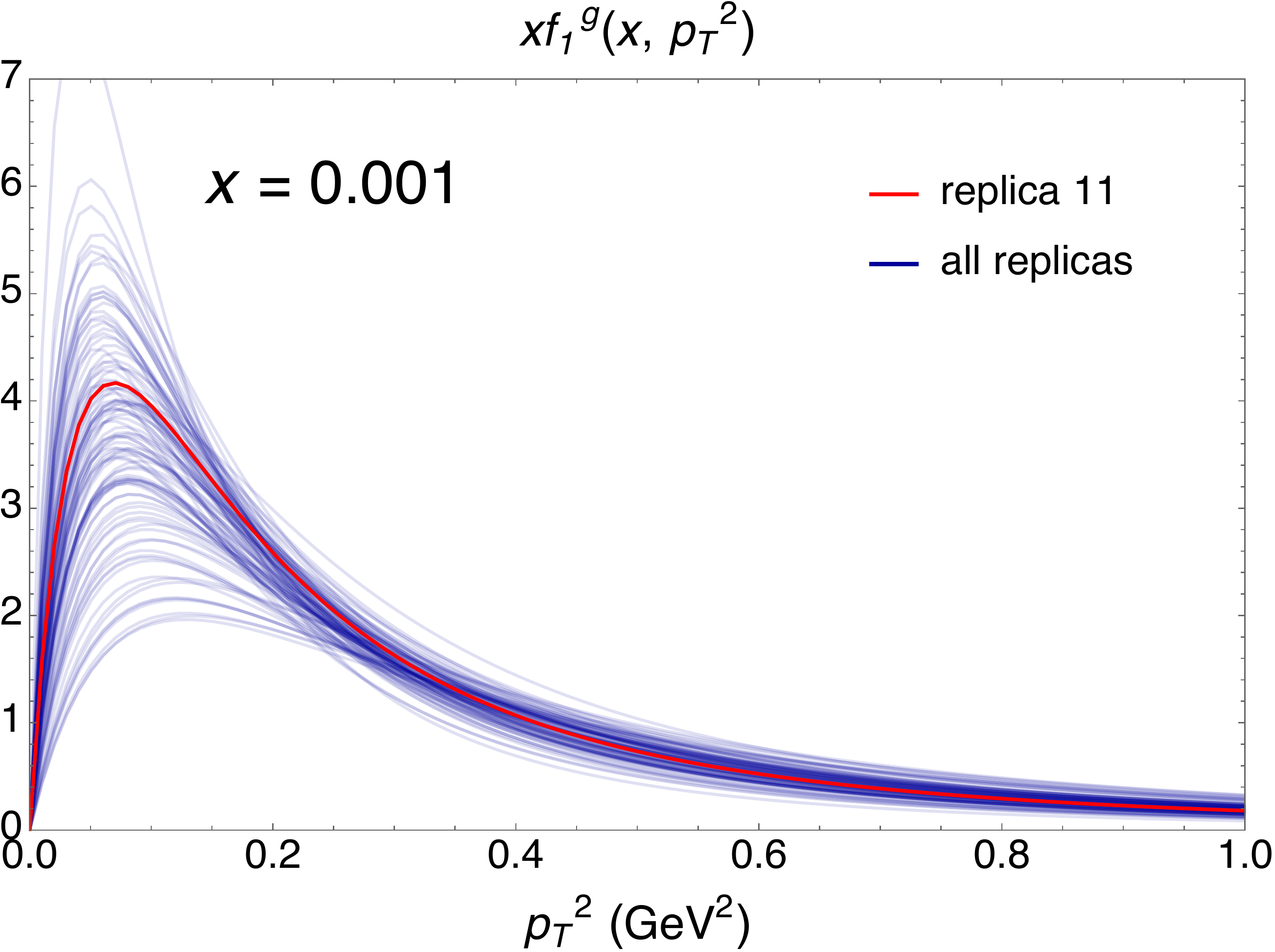}

 \vspace{0.50cm}

 \includegraphics[width=0.48\textwidth]{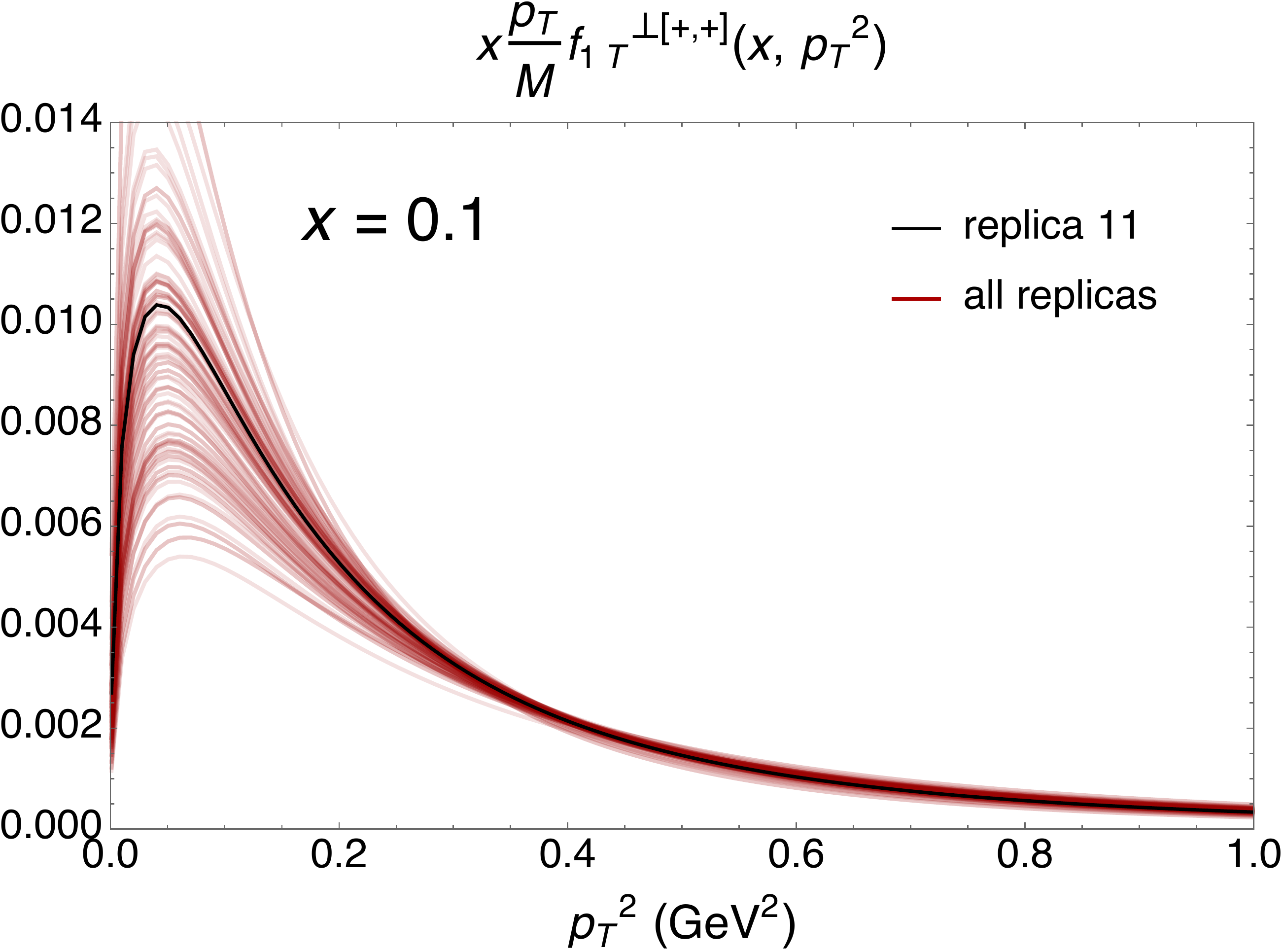}
 \hspace{0.25cm}
 \includegraphics[width=0.48\textwidth]{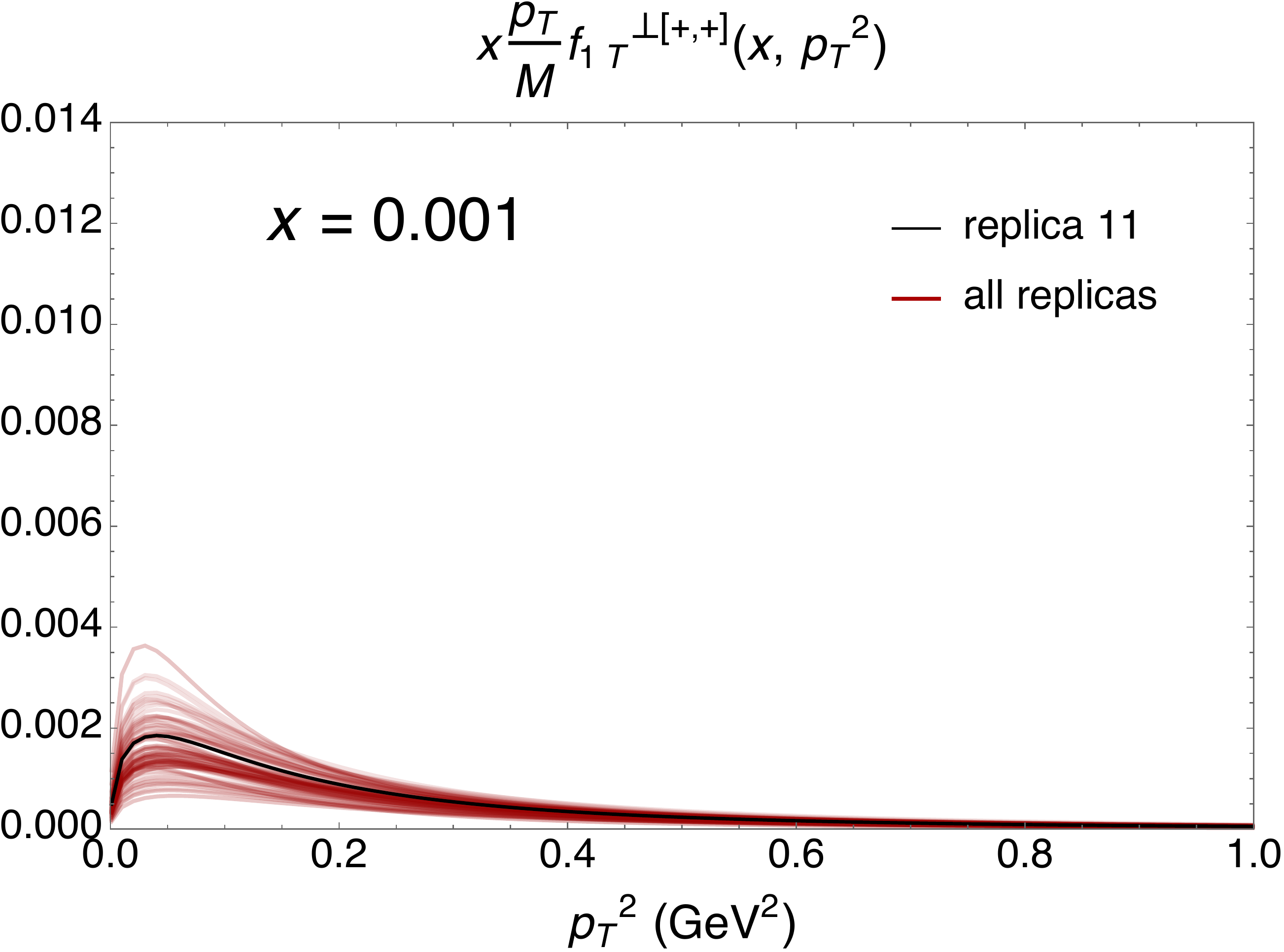}

 \caption{Transverse-momentum dependence of the $[+,+]$ Sivers (upper) and linearity (lower) densities for $x=10^{-3}$ (left) and $x=10^{-1}$ (right), and at the initial scale $Q_0 = 1.64$ GeV. Black curves refer to the most representative replica \#11.}
\label{fig:TMD_PDFs}
\end{figure*}

Accessing the inner structure of hadrons via a multi-dimensional study of their constituents represents a frontier research of phenomenological analyses at new-generation accelerators.

In the last decades the well-established collinear factorization based on a one-dimensional description of the proton content through collinear parton density functions (PDFs) has obtained a long series of successes in describing data at hadron and lepton-hadron colliders.

On the other hand, essential questions on the dynamics of strong interactions still do not have an answer.
Unraveling the origin of mass and spin of the nucleons calls for an enlargement of our point of view from the purely collinear picture to a tomographic vision in three dimensions, naturally afforded by the transverse-momentum-dependent (TMD) formalism.

Our knowledge of gluon TMDs is much more limited than the one of quark TMDs.
In~\cite{Mulders:2000sh} and then in~\cite{Meissner:2007rx,Lorce:2013pza,Boer:2016xqr} (un)polarized TMD densities were classified for the first time, while first attempt at phenomenological analyses can be found in~\cite{Lu:2016vqu,Lansberg:2017dzg,Gutierrez-Reyes:2019rug,Scarpa:2019fol,COMPASS:2017ezz,DAlesio:2017rzj,DAlesio:2018rnv,DAlesio:2019qpk}.

One of the major distinction between collinear and TMD distributions is their dependence on gauge links.
More in particular, the sensitivity of TMDs to the transverse part of the gauge link leads to a process dependence which is instead not present in the collinear case~(see Refs.~\cite{Brodsky:2002cx,Collins:2002kn,Ji:2002aa}).
The process dependence of quark TMDs is realized via the $[+]$ and $[-]$ staple links, which are respectively connected the future- and the past-pointing direction of Wilson lines.
Gluon TMDs exhibit a more intricate gauge-link dependence, which brings to a more complicated \emph{modified universality}. This comes out from their sensitivity to a combination of two or more staple links.
There are two main gluon gauge-link structures: the $f\text{-type}$ and the $d\text{-type}$ one, respectively known in the context of small-$x$ studies as Weisz\"acker--Williams and dipole links.
The antisymmetric QCD color structure, $f_{abc}$, is part of the $f$-type $T$-odd gluon-TMD correlator, while the $d$-type $T$-odd one contains the symmetric color structure, $d_{abc}$.
The $f$-type gluon TMDs are sensitive to the $[\pm,\pm]$ gauge-link combinations, while $d$-type gluon TMDs are characterized by the $[\pm,\mp]$ ones.
More complicated gauge-link structures, given in terms of box-loop combinations of $[+]$ and $[-]$ staple links, are probed via reactions featuring multiple color exchanges between the initial and the final state states~\cite{Bomhof:2006dp}. For these processes, however, it was proven that the TMD factorization is violated~\cite{Rogers:2013zha}.

In the small-$x$ and moderate-$p_T$ regime the unpolarized gluon TMD is connected with the BFKL unintegrated distribution~\cite{Dominguez:2011wm,Nefedov:2021vvy,Hentschinski:2021lsh}, of which 
several recent studies have appeared so far~\cite{Bolognino:2018rhb,Bolognino:2018mlw,Bolognino:2019bko,Bolognino:2019pba,Celiberto:2019slj,Celiberto:2018muu,Bautista:2016xnp,Hentschinski:2020yfm}. Analyses on the dynamics of strong interactions at high energies via the BFKL approach were recently performed in~\cite{Celiberto:2020wpk,Celiberto:2016vhn,Bolognino:2018oth,Bolognino:2019yqj,Bolognino:2019yls,Bolognino:2021niq,Celiberto:2020tmb,Celiberto:2020rxb,Bolognino:2021mrc,Celiberto:2021dzy,Celiberto:2021fdp,Celiberto:2021txb,Celiberto:2021xpm,Bolognino:2021zco,Celiberto:2021tky,Celiberto:2021fjf,Bolognino:2021hxx,Celiberto:2022dyf,Celiberto:2022rfj}.

A study of the proton content via quark TMDs calculated in the spectator-model formalism was done in~\cite{Bacchetta:2008af,Bacchetta:2010si}, by accounting for different polarization states of the di-quark spectators and various form factors for the nucleon-parton-spectator vertex.
A common formalism was recently set up~\cite{Bacchetta:2020vty} (see also~\cite{Bacchetta:2021oht,Celiberto:2021zww,Bacchetta:2021lvw,Bacchetta:2021twk,Celiberto:2022fam}) for all the leading-twist $T$-even gluon TMD densities in the proton. Here, the standard spectator model was improved to effectively capture effects coming from the high-energy dynamics of QCD.

In this work we present results for the $T$-even unpolarized gluon TMD and a preliminary study on the $f$-type $T$-odd gluon Sivers distribution, both of them calculated in the spectator-model approximation.

\section{Gluon TMD distributions in a spectator model}
\label{gluon_TMDs}

In the spectator framework one models the gluon correlator in the following way. From an incoming nucleon with mass $\cal M$ and four-momentum $\cal P$ a gluon is emitted with longitudinal fraction $x$, four-momentum $p$, and transverse momentum $\boldsymbol{p}_T$. What remains is effectively described as a single on-shell spectator having mass ${\cal M}_X$ and spin-1/2.
The model for the nucleon-gluon-spectator vertex is the following
\begin{equation}
 \label{eq:form_factor}
 {\Gamma}^{\, \mu} = \left( \Xi_1(p^2) \, \gamma^{\, \mu} + \Xi_2(p^2) \, \frac{i}{2{\cal M}} \sigma^{\, \mu\nu}p_\nu \right) \; ,
\end{equation}
the $\Xi_1$ and $\Xi_2$ functions taken as dipolar coupling in the gluon transverse momenta. Having dipolar form factors permits to dampen gluon-propagator divergences, quench large-$\boldsymbol{p}_T$ effects, and remove logarithmic singularities arising in $\boldsymbol{p}_T$-integrated densities.
Taking into account the corresponding nucleon and parton polarization states, we calculated all the $T$-even gluon TMD distributions at leading twist in the proton in the spectator-model approach\tcite{Bacchetta:2020vty}. 
In that paper the naive spectator formalism was enhanced by weighting the mass of the spectator ${\cal M}_X$ over a continuous range though a spectral function suited to catch both small- and moderate-$x$ effects.
The values of model parameters embodied in the spectral mass and in the nucleon-gluon-spectator vertex were fixed via a simultaneous fitting procedure of the $|\boldsymbol{p}_T|$-integrated unpolarized and helicity gluon TMDs, $f_1^g$ and $g_1^g$, to the corresponding collinear PDF distributions obtained by the {\tt NNPDF} collaboration\tcite{Ball:2017otu,Nocera:2014gqa} at the initial scale of $Q_0 = 1.64$ GeV. The impact of the statistical uncertainty was evaluated via the well-known bootstrap method.

The gluon correlator at tree-level does not account for the gauge link. Therefore, our $T$-even TMD densities do not exhibit any process dependence.
$T$-odd structures can be generated by going beyond the tree-level approximation for the gluon correlator, and considering its interference with a distinct channel. In the same way as quark TMDs\tcite{Bacchetta:2008af}, we have accounted for the one-gluon exchange in the \emph{eikonal} approximation, which corresponds to the first-order truncation of the complete gauge-link operator. As a major effect, the obtained $T$-odd densities become sensitive to gauge links, and thus they depend on the process. For a given gauge link, two Sivers TMDs ($f_{1T}^\perp$) exist and are obtained after a suitable projection of the transverse component of the gluon correlator. The following relations of modified universality hold
\begin{align}
 \label{eq:T-odd_TMD_PDFs}
 f_{1T}^{\perp \, [+,+]}(x, \boldsymbol{p}_T^2) \; & = \; - \, f_{1T}^{\perp \, [-,-]}(x, \boldsymbol{p}_T^2) \; ,
 \\ \nonumber
 f_{1T}^{\perp \, [+,-]}(x, \boldsymbol{p}_T^2) \; & = \; - \, f_{1T}^{\perp \, [-,+]}(x, \boldsymbol{p}_T^2) \; .
\end{align}
We made a preliminary study of the $[+,+]$ Sivers function with a simplified formula for the nucleon-gluon-spectator vertex, obtained by setting the $\Xi_2$ form factor in Eq.\tref{eq:form_factor} to zero.
Here, by reason of consistency, we fitted again the model parameters to {\tt NNPDF} parameterizations by adopting the simplified formula of the vertex.

Upper panels of Fig.\tref{fig:TMD_PDFs} show the dependence on transverse momentum of the unpolarized gluon TMD for two values of the longitudinal-momentum fraction, $x = 10^{-1}$ and $x = 10^{-3}$, and with the initial scale fixed to $Q_0 = 1.64$ GeV. Lower panels contain the dependence of the $p_T$-weighted $[+,+]$ Sivers distribution for the same values of $x$ and $Q_0$. Both the densities exhibit a non-Gaussian $\boldsymbol{p}_T^2$-behavior, with a flattening tail at large transverse momenta and a small but nonzero value when $\boldsymbol{p}_T^2 \to 0$. This suggests that in the low-$\boldsymbol{p}_T$ limit both functions diverge at most as $1/|\boldsymbol{p}_T|$. 
The two densities exhibit an opposite behavior in $x$. While the unpolarized TMD has a bulk that grows when $x$ diminishes, the peak of the Sivers functions shrinks at low $x$. This is an indication that transverse single-spin asymmetries are expected to be less significant in the low-$x$ range.
However, we stress that results obtained for the Sivers density could change even drastically when the full-vertex calculation will be afforded.

\section{Conclusions}
\label{summary}

We calculated all leading-twist $T$-even gluon TMDs by the hands of an enhanced spectator-model framework, which allowed us to catch both the small- and the moderate-$x$ range. The full calculation of the leading-twist $f$-type $T$-odd TMDs, such as the Sivers function is underway. They can be used as a useful guidance to unveil the gluon-TMD dynamics at new-generation colliding facilities, such as the Electron-Ion Collider~(EIC)~\cite{AbdulKhalek:2021gbh}, NICA-SPD~\cite{Arbuzov:2020cqg}, the \emph{High-Luminosity Large Hadron Collider} (HL-LHC)~\cite{Chapon:2020heu}, and the \emph{Forward Physics Facility} (FPF)~\cite{Anchordoqui:2021ghd,Feng:2022inv}.

\end{multicols}
\medline
\begin{multicols}{2}

\end{multicols}

\end{document}